# HELEN: A LINEAR COLLIDER BASED ON ADVANCED SRF TECHNOLOGY*

S. Belomestnykh[†,1], P. C. Bhat, M. Checchin[‡], A. Grassellino, M. Martinello[‡], S. Nagaitsev[2],
H. Padamsee[3], S. Posen, A. Romanenko, V. Shiltsev, A. Valishev, V. Yakovlev
Fermi National Accelerator Laboratory, Batavia, IL, USA
[1]also at Stony Brook University, Stony Brook, NY, USA
[2]also at University of Chicago, Chicago, IL, USA
[3]also at Cornell University, Ithaca, NY, USA

## Abstract

This paper discusses recently proposed Higgs-Energy LEptoN (HELEN) $e^+e^-$ linear collider based on advances in superconducting radio frequency technology. The collider offers cost and AC power savings, smaller footprint (relative to the ILC), and could be built at Fermilab with an interaction region within the site boundaries. After the initial physics run at 250 GeV, the collider could be upgraded either to higher luminosity or to higher (up to 500 GeV) energies.

## INTRODUCTION

One of the highest priorities for the particle physics community is to make precision measurements of the Higgs boson properties and look for any deviations from the Standard Model using an $e^+e^-$ collider at the center-of-mass energy of 250 to 360 GeV (Higgs factory). For many years, the International Linear Collider (ILC) has been the prime candidate for such a machine. Its mature superconducting radio frequency (SRF) technology has been "shovel ready" and has been used already to build such linacs as European XFEL in Hamburg, Germany, and LCLS-II at SLAC in the USA. Meanwhile, the SRF community continues to make progress improving the performance of SRF cavities.

In this paper we discuss how recent advances in the SRF technology can be applied for a more compact and cost-effective $e^+e^-$ linear collider. This recently proposed machine is named Higgs-Energy LEptoN (HELEN) collider [1]. If the ILC cannot be realized in Japan in a timely manner, HELEN could be built after relatively short period of dedicated R&D efforts. At the core of this collider is the traveling wave (TW) SRF technology. The paper describes the machine in some detail including tentative list of parameters, layout and possible siting, and potentials for luminosity and energy upgrades. Finally, we provide summary and conclusions.

## PROMISE OF TRAVELING WAVE SRF

Travelling wave structures offer several advantages over the traditional standing wave SRF structures: substantially lower $H_{pk}/E_{acc}$ and lower $E_{pk}/E_{acc}$, ratios of peak magnetic field and peak electric field to the accelerating gradient, respectively, together with substantially higher $R/Q$. To reach the maximal gradient, the optimal TW cavity design must have the lowest possible $H_{pk}/E_{acc}$, since $H_{pk}$ presents the hard ultimate limit to the performance of Nb cavities via RF superheating field.

On the other hand, as Fig. 1 shows, the TW structure requires almost twice the number of cells per meter compared to the SW structure to provide the proper phase advance (105° in Fig. 1), as well as a feedback waveguide for redirecting power from the end to the front of the accelerating structure. The feedback waveguide requires careful tuning to compensate reflections along the TW ring and thus obtain a pure traveling wave regime at the desired frequency.

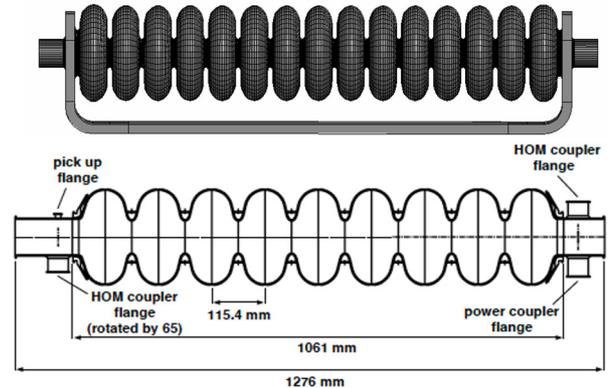

Figure 1: A one-meter-long TW structure with a 105° phase advance per cell compared to the one-meter standing-wave TESLA structure [2]. Note that a TW structure for HELEN will be longer, perhaps 2-meter long.

Recent TW cavity geometry optimization study [3] demonstrated that for an aperture radius $R_a = 25$ mm and phase advance of 90°, one can achieve $H_{pk}/E_{acc} = 28.8$ Oe/(MV/m) with $E_{pk}/E_{acc} = 1.73$. Since $H_{pk}/E_{acc}$ is 42.6 Oe/(MV/m) for the TESLA structure, the TW structure has reduced the critical parameter $H_{pk}/E_{acc}$ by almost a factor of 1.5. At the same time, the peak electric field ratio is smaller than TESLA cavity value of 2.0 and we gain a factor of 2.1 in $R/Q$ although losing in the geometry factor by 1.45 (186 Ohm vs. 270 Ohm). Thus, for the same accelerating gradient the TW cavity is more cryogenically efficient than TESLA cavity. The high group velocity in the TW mode increases the cell-to-cell coupling from 1.8% for the TESLA structure to 2.3%. Thus, TW structures have less sensitivity to cavity detuning errors, making tuning easier, despite the larger number of cells. The high stability of the field distribution along the structure with respect to geometrical perturbations allows for much longer (limited by manufacturing technology) accelerating structures than TESLA cavities. We anticipate that 2-meter long traveling wave accelerating structures would be feasible, resulting in fewer cavity-to cavity transitions in the linac and hence to larger real-estate gradients.

___________
* Work supported by the Fermi National Accelerator Laboratory, managed and operated by Fermi Research Alliance, LLC under Contract No. DE-AC02-07CH11359 with the U.S. Department of Energy.
† sbelomes@fnal.gov
‡ present affiliation: SLAC, Menlo Park, CA, USA

Furthermore, the cell shape can be fine-tuned to avoid multipacting without increasing $H_{pk}$ more than 1%. While higher order mode (HOM) damping for TW structures is under study, preliminary results indicate that the first 10 monopole modes up to 7 GHz show no trapping [3].

What accelerating gradient can we expect to reach with traveling wave structures? A new two-step low temperature (75/120°C) baking in combination with cold electro-polishing pushes quench field of 1.3 GHz single-cell TESLA shape cavities to 50 MV/m [4, 5]. This gradient corresponds to the peak surface magnetic field of 2130 Oe. Assuming that after applying similar cavity treatment the TW cavity can demonstrate similar magnetic quench limit, we can expect that TW cavities should be capable of reaching 74 MV/m.

## HELEN COLLIDER

In his paper [6], Padamsee considered several options for ILC energy upgrade. Rather than waiting for ILC upgrades, it was proposed to use these advanced technological options from the very beginning as potential technologies for a compact $e^+e^-$ linear collider HELEN [1]. Here we consider only the baseline SRF technology option: the traveling wave SRF linac. This option provides an optimal combination of the high accelerating gradient of 70 MV/m with an expected demonstration of a fully developed cryomodule within ~5 years if sufficient funding for the R&D is available. Besides the SRF structure, HELEN is very similar to ILC, and the collider layout (shown in Fig. 2) and most of its parameters (listed in Table 1) are similar to ILC as well. HELEN's AC power consumption is similar to the ILC power demand. However, due to higher accelerating gradient (and larger fill factor of 80.4% vs. 71% for ILC), it offers significant main linac cost saving of about 37%. All proposed ILC luminosity upgrade scenarios (see, e.g., [7]) are applicable to HELEN.

We have identified locations for a possible future linear collider at Fermilab [8]: two 7-km diagonal options and a 12-km footprint with N-S orientation extending outside the site boundary but with the Interaction Region (IR) on site (Fig. 3). A 250-GeV HELEN Higgs factory (HELEN-250) could potentially fit along either of two diagonals after further optimization of the collider. The 12-km N-S footprint can accommodate not only the 7.5-km-long (including 3 km of beam delivery system) HELEN-250 but would also allow extension of the main linacs to the center-of-mass energy of 500 GeV (HELEN-500).

Table 1: Tentative Baseline Parameters of HELEN

| Parameter | Value |
| --- | --- |
| Center of mass energy | 250 GeV |
| Collider length | 7.5 km |
| Peak luminosity | $1.35 \times 10^{34}$ cm$^{-2}$s$^{-1}$ |
| Repetition rate | 5 Hz |
| Bunch spacing | 554 ns |
| Particles per bunch | $2 \times 10^{10}$ |
| Bunches per pulse | 1312 |
| Pulse duration | 727 μs |
| Pulse beam current | 5.8 mA |
| Bunch length, rms | 0.3 mm |
| Crossing angle | 14 mrad |
| Crossing scheme | crab crossing |
| RF frequency | 1300 MHz |
| Accelerating gradient | 70 MV/m |
| Real estate gradient | 55.6 MV/m |
| Total site power | 110 MW |

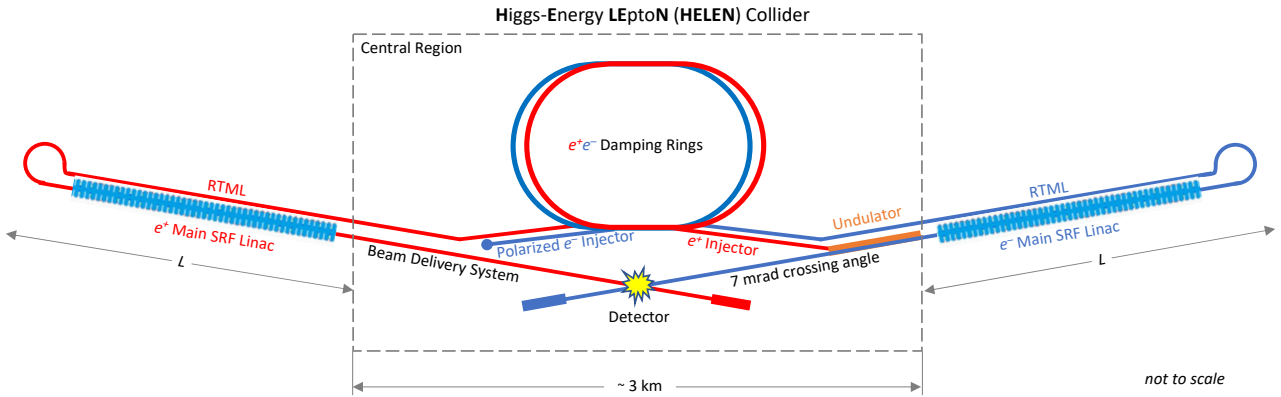

Figure 2: Conceptual layout of the HELEN collider.

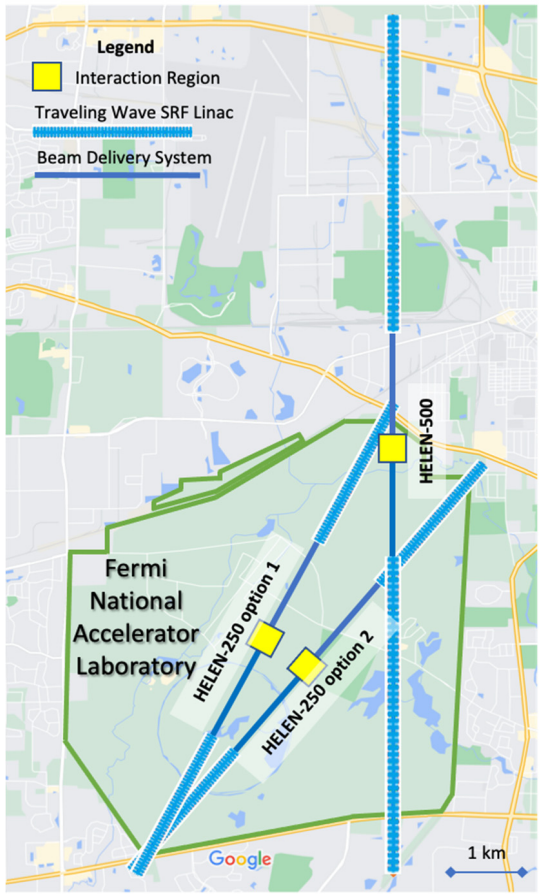

Figure 3: Options for HELEN collider at Fermilab.

## HELEN ACCELERATOR R&D PROGRAM

The major objective of the accelerator R&D program is to demonstrate feasibility of the traveling wave technology:

- Test proof-of-principle multi-cell 1.3 GHz TW cavity and demonstrate accelerating gradient of ~70 MV/m.
- Adapt an advanced cavity treatment technique so that high Q ~ $10^{10}$ can be achieved at high gradient.
- Design, build and test full-scale prototype cavities; demonstrate performance needed for the HELEN collider.
- Design and build a prototype cryomodule for TW SRF cavities.
- Verify the cryomodule performance without beam on a test stand and with beam at Fermilab's FAST facility.

Demonstration efforts have started with a single-cell TW cavity development [9]. The first cavity with recirculating waveguide achieved 26 MV/m accelerating gradient, limited by the high field Q-slope, as expected for BCP treatment. This result was very encouraging for the first attempt. This was followed by development of a 3-cell Nb TW structure. The cavity was designed and fabricated (see Fig. 4) [10]. Currently, it is undergoing tuning and treatment at Fermilab with the first vertical test scheduled for fall 2022.

Beyond the demonstration of TW technology, the R&D program will pursue the following tasks:

- Design and optimization of the HELEN linear collider accelerator complex.
- Confirm the physics reach and detector performance for the HELEN beam parameters.
- Publish Conceptual Design Report as modification of the ILC design in 2–3 years.
- Prepare Technical Design Report after demonstrating the cryomodule performance, in ~5 years.

## SUMMARY AND CONCLUSIONS

If the ILC project in Japan will not gain traction, the expertise and technological advances accumulated by the world SRF community – in particular, at the U.S. laboratories and universities (Cornell, Fermilab, JLAB, SLAC, ...) – would allow rapid development, prototyping, and testing of SRF cavities and cryomodules based on the traveling wave technology for a new, compact $e^+e^-$ SRF linear collider HELEN. Fermilab has capabilities that support the full cycle of R&D, production, and verification (including testing cryomodules with beam) at the SRF accelerator test facilities and FAST linac.

If given high priority, the construction of the 250-GeV center of mass energy HELEN collider could start as early as 2031–2032 with first physics in ~2040. The HELEN collider can be upgraded to higher luminosities in the same way as was proposed for the ILC or to higher energies (up to 500 GeV at Fermilab site) by extending the linacs.

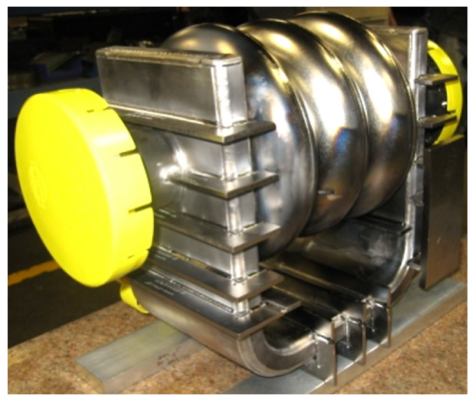

Figure 4: Fabricated 3-cell traveling wave cavity [10].